# The Network Species Model

DAVID KIZIRIAN AND MAUREEN A. DONNELLY

*Biological Sciences, Florida International University, Miami, FL 33199*

**ABSTRACT:** We propose a theoretical model for species that is focused on intrinsic organization and processes. Specifically, we regard species to be networks of organisms integrated by reproductive mechanisms (e.g., conjugation, meiosis, syngamy) regardless of degree or type of divergence, longevity, size, fate, or other criteria. Ramifications of viewing units of diversity in terms of inherent organization include the following: (1) Species, subspecies, population, deme, Evolutionary Significant Unit (ESU) and related terms cannot be distinguished on the basis of inherent organization and, consequently, do not reflect a hierarchy of organization. In other words, only one model is required to explain systems composed of organisms. (2) Temporarily isolated networks of organisms possess the same intrinsic organization and, therefore, the same ontological status as permanently isolated systems. (3) Units of diversity are recognized regardless of kind or degree of divergence. (4) Fate is not an intrinsic property of systems and is an inappropriate consideration in species models. (5) Networks are recognized at the moment they become isolated; therefore, resultant classifications reflect the causal events that generate diversity (e.g., vicariance) rather than post-vicariance events such as character evolution and, consequently, have greater historical relevance (e.g., biogeography). (6) Because the proposed model is based on organization, it is similar to models in other disciplines (e.g., astronomy, chemistry, and physics) and should lead to greater unification of thought within biology and among scientific disciplines. (7) Lineage concepts are problematic because they do not describe a unique level of biological organization and because ancestor-descendant relationships are not unique to living systems.

\*                                                          \*                                                          \*

## Introduction

We explore the idea that theory about species can be improved by focusing on the organization and processes unique to systems composed of organisms. We argue that the term *species* is best associated with a highly simplified model, i.e., *networks of organisms that form through reproduction* (in italics here and throughout). The proposed model highlights organization (i.e. network of organisms) and processes that maintain such organization (e.g., conjugation, meiosis, syngamy). Furthermore, the model eschews properties characteristic of many species concepts. In other words, networks—regardless of degree or type of divergence, longevity, size, or fate—are the fundamental units of study for biodiversity research.  For example, we argue that the Mole Salamander (*Ambystoma talpoideum*) is a complex of at least 10 species (Fig. 1). Recognition of the isolated networks of *A. talpoideum* as species reflects the structure of systems (i.e., networks of organisms) rather the criteria associated with other models such as fate or divergence (e.g., intrinsic reproductive isolation). In addition, under the proposed model, a breeding colony of a *Drosophila* in a test tube has the same ontological status as the Cuban Crocodile (*Crocodylus rhombifer*), currently known only from a single swamp in Cuba, because both have the same organization (i.e., they are networks of organisms).
\*\*\*\*\*\*\*\*\*\*\*\*\*\*\*\*\*\*\*\*\*\*\*\*\*\*\*\*\*\*\*\*\*\*\*\*\*\*\*\*\*\*\*\*\*\*\*\*\*\*\*\*\*\*\*\*\*\*\*\*\*\*\*\*\*\*\*\*\*\*\*\*\*\*\*\*\*\*\*\*\*\*\*\*\*\*\*\*\*\*\*\*\*\*\*\*\*\*\*\*\*\*\*\*\*\*\*\*\*\*\*\*\*

Figure 1: Distribution map for the Mole Salamander, *Ambystoma talpoideum* (from Conant and Collins 1991). Under current concepts of species, isolated networks are regarded to be populations (one level of organization) of a single species (another level of organization). Under the proposed model, at least ten species (and, therefore, one level of organization) would be recognized, which reflects system structure rather than reproductive compatibility, degree of divergence, or fate of isolated networks. Classifications consistent with the proposed species model include (1) "*Ambystoma talpoideum* complex" and (2) unique binominals (appended, or not, with "complex" as appropriate) for each isolated network.

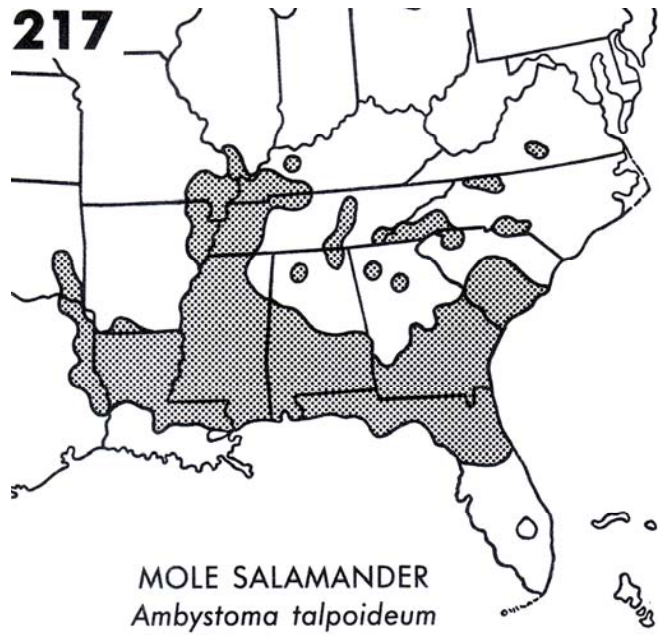

MOLE SALAMANDER
*Ambystoma talpoideum*

\*\*\*\*\*\*\*\*\*\*\*\*\*\*\*\*\*\*\*\*\*\*\*\*\*\*\*\*\*\*\*\*\*\*\*\*\*\*\*\*\*\*\*\*\*\*\*\*\*\*\*\*\*\*\*\*\*\*\*\*\*\*\*\*\*\*\*\*\*\*\*\*\*\*\*\*\*\*\*\*\*\*\*\*\*\*\*\*\*\*\*\*\*\*\*\*\*\*\*\*\*\*\*\*\*\*

The following extend from focus on the inherent organization of systems composed of organisms and the processes that maintain such systems. (1) Species, subspecies, population, deme, Evolutionary Significant Unit (ESU), Management Unit (MU), Distinct Population Segment (DPS) and related terms (e.g., Du Rietz, 1930) cannot be distinguished on the basis of inherent organization and, consequently, do not reflect a hierarchy of organization. In other words, only one model is required to explain systems composed of organisms. (2) Temporarily isolated networks of organisms possess the same intrinsic organization and, therefore, the same ontological status as permanently isolated systems. (3) Units of diversity are recognized regardless of kind or degree of divergence. (4) Fate is not an intrinsic property of systems and is an inappropriate consideration in species models. (5) Networks are recognized at the moment they become isolated; therefore, resultant classifications reflect the causal events that generate diversity (e.g., vicariance) rather than post-vicariance events such as character evolution and, consequently, have greater historical relevance (e.g., biogeography). (6) Because the proposed model is based on organization, it is similar to models in other disciplines (e.g., astronomy, chemistry, and physics) and should lead to greater unification of thought within biology and among scientific disciplines. (7) Lineage concepts are problematic because they do not describe a unique level of biological organization and because ancestor-descendant relationships are not unique to living systems.

One source of obfuscation regarding the term species stems from its usage in conjunction with taxonomic categories (e.g., phyla, classes, orders, families), which are widely acknowledged to be subjective and biologically meaningless (Ereshefsky, 1997, 2002, 2007; Minelli, 2000; Laurin, 2005). Here, we do not address at length operational matters, species as taxa, taxonomy, or rules of nomenclature. Rather, we focus our discussion on species as systems in nature and attempt to articulate a model that best explains such systems. It is relevant to note, however, that existing nomenclatural protocols can accommodate the great diversity implied by the proposed model. For example, "species complex" nomenclature (e.g., *Gasterosteus aculeatus* complex; e.g., Schluter, 1996) has long been used to recognize abundant or ambiguous species-level diversity. We are not opposed to viewing *species* as an arbitrarily defined rank in a taxonomic hierarchy as long as the term is not used differently from other ranks. Nor are we opposed to jettisoning the term in favor of one with less baggage. As long as *species* implies that the thing being named is ontologically unique with respect to other named things (as is the case with most species concepts), it is in the best interest of science to improve the theoretical underpinnings of the term.

Some points regarding our approach warrant mention. We do not provide an exhaustive review of species models (e.g., Mayden, 1997), rather, we propose a new way of looking at diversity and note some of the advantageous ramifications of that proposal. We use language that is widely appreciated (Oxford English Dictionary [OED], 2006) and minimize use of terms from the philosophical literature. Departing from standard practice, we propose a *model* rather than a *concept* because the former tend to be more explicit and focused on structure and the latter tend to be more nebulous (OED, 2006). Where appropriate, we cite a dictionary (OED, 2006) and an introductory biology text (Raven et al. 2002) because we think species models should be accessible at those levels of understanding, as is the case for other levels of organization. We cite Raven *et al.* (2002) in particular because most of those authors have been involved in biodiversity research and presumably should be able to provide clear distinctions among relevant terms in that discipline.

## Historical review

It has been argued that much diversity and complexity in the natural world is explained by relatively few models (e.g., Bonner and Collins, 1969; Griffiths, 1974; Hey, 2001; Raven *et al.*, 2002; Tables 1–2). Additional complexity can be explained by the hierarchical relationships among some models (e.g., Burma, 1954; Griffiths, 1974; Vrba and Eldredge, 1984; Eldredge, 1985; Salthe, 1985; Grene, 1987; Mahner and Bunge, 1997; Ghiselin, 1992b; Zylstra, 1992; Salthe, 1993; Frost and Kluge, 1994; Ghiselin, 1997; Raven *et al.*, 2002; Tables 1–2). Species are generally thought to represent a particularly significant level in the hierarchy of biological organization (e.g., Du Rietz, 1930; Mayr, 1957a; Grant, 1971; Griffiths, 1974; Mayr, 1976; Ghiselin, 1997:12; Cohan, 2002; De Queiroz, 2005). Ironically, however, organization has not been emphasized in species concepts, particularly since the Modern Synthesis. Highlighted instead are contingent properties (e.g., De Queiroz, 1998), including properties that emerge from the inherent organization of species (such as divergence or other criteria), rather than the organizational aspects themselves. Because complex systems such as species typically exhibit a diversity of properties, subjective evaluation of the significance of those properties necessarily precedes their inclusion or exclusion in species models. Such subjectivity may partially explain the longevity of the species concept debate (De Queiroz, 1998) and is likely to be manifested in a misunderstanding of biological diversity (see Jonckers, 1973 for a similar discussion regarding the term population).

**************************************************************************************************************************

Table 1: Widely available terms for some levels of organization relevant to biology (*ex* Oxford English Dictionary, 2006). Terms without an asterisk are characterized by unique inherent structure and terms with an asterisk are not. We argue that the ambiguity in terms available to the general public stems from poor modeling by biologists (e.g., Raven et al., 2002).

Atom: A positively charged nucleus, in which is concentrated most of the mass of the atom, and round which orbit negatively charged electrons.

Molecule: The smallest unit of a chemical compound that can take part in the reactions characteristic of that compound; a group of atoms chemically bonded together and acting as a unit.

Cell: The ultimate element in organic structures; a minute portion of protoplasm, enclosed usually in a membranous investment.

Tissue: Any one of various structures, each consisting of an aggregation of similar cells or modifications of cells which make up the organism.

Organ: A part or member of an animal or plant body adapted by its structure for a particular vital function, as digestion, respiration, excretion, reproduction, locomotion, perception, etc.

Organism: An organized body, consisting of mutually connected and dependent parts constituted to share a common life; the material structure of an individual animal or plant.

*Deme: A local population of closely related plants or animals.

*Population: A breeding group of animals, plants, or humans.

*Subspecies: A subdivision of a species; a more or less permanent variety of a species.

*Species: A group or class of animals or plants (usually constituting a subdivision of a genus) having certain common and permanent characteristics which clearly distinguish it from other groups.

Ecosystem: The unit of ecology is the ecosystem, which includes the plants and animals occurring together plus that part of their environment over which they have an influence.

**************************************************************************************************************************

In contrast, models for other levels of organization, such as atom, molecule, and cell, are less problematic and more widely appreciated (e.g., Oxford English Dictionary [OED] 2006; Table 1). The success of such models may stem from their emphasis on the more objectively evaluated fundamental organization of the systems they describe, rather than more subjectively evaluated contingent properties. For example, one widely accepted model for atom (OED 2006; Table 1), which is a simple description of its parts and how they are integrated, possesses enormous explanatory power and is characterized by objectivity and clarity for scientists and non-scientists. Absent from this model are properties that extend from the inherent organization of atoms such as the tendency to form ionic or covalent bonds.

\*\*\*\*\*\*\*\*\*\*\*\*\*\*\*\*\*\*\*\*\*\*\*\*\*\*\*\*\*\*\*\*\*\*\*\*\*\*\*\*\*\*\*\*\*\*\*\*\*\*\*\*\*\*\*\*\*\*\*\*\*\*\*\*\*\*\*\*\*\*\*\*\*\*\*\*\*\*\*\*\*\*\*\*\*\*\*\*\*\*\*\*\*\*\*\*\*\*\*\*\*\*\*\*\*\*\*\*\*\*\*\*\*\*\*\*\*\*\*\*\*\*\*

Table 2:  Examples of some hierarchical and non-hierarchical models of organization. Non-hierarchical models are also examples of corresponding hierarchical models. Examples of historical individuals are also provided for some categories.

| Hierarchical | Non-hierarchical | Individuals |
|---|---|---|
| Ecosystem | grassland | Konza Prairie |
| Species | syngamic system | *Crocodylus rhombifer* |
| Organism | multicellular | Frank Zappa |
| Organ system | digestive system | — |
| Organ | liver | Mickey Mantle's first liver |
| Tissue | striated muscle | — |
| Cell | prokaryote | — |
| Molecule | double helix | — |
| Atom | metal | — |

\*\*\*\*\*\*\*\*\*\*\*\*\*\*\*\*\*\*\*\*\*\*\*\*\*\*\*\*\*\*\*\*\*\*\*\*\*\*\*\*\*\*\*\*\*\*\*\*\*\*\*\*\*\*\*\*\*\*\*\*\*\*\*\*\*\*\*\*\*\*\*\*\*\*\*\*\*\*\*\*\*\*\*\*\*\*\*\*\*\*\*\*\*\*\*\*\*\*\*\*\*\*\*\*\*\*\*\*\*\*\*\*\*\*\*\*\*\*\*\*\*\*\*

Systems characterized by unique organization also generally exhibit unique intrinsic *processes* (see also Griffiths, 1974). For example, protein synthesis occurs only in cells, nutrient cycling occurs only in ecosystems, and nuclear fusion occurs naturally only in stars. We interpret the existence of unique processes to be an indicator of the existence of unique system organization. In particular, we observe that conjugation, meiosis, and syngamy are among the unique processes that integrate networks of organisms. The species model we promote draws attention to the systems where such processes occur.

Early species models, including those predating Darwin (1859), tended to focus on intrinsic organization and process, at least when interpreted literally. For example, the models of Illiger (1800; "a community of individuals which produce fertile offspring") and Voigt (1817; "Whatever interbreeds fertilely and reproduces is called a species") are relatively simple and emphasize the structure of species, i.e., a network of individual organisms connected by mating events. Poulton (1904), an early evolutionary biologist who influenced Mayr (e.g., Mallet, 2004), proposed a model that explicitly describes the nature of the relationships among individual organisms in such networks, i.e., "societies into which individuals are bound together in space and time by syngamy and epigony" or "syngamic communities" (see also Mallet, 2004; epigony refers to the lineage relationship that can be observed by an investigator, e.g., matings and subsequent hatchings). Naef (1919: 44–46) proffered a similar model:

> This forces us to focus more closely on the species model, as it appears in the basic notion of the theory of descent: In nature first individuals are given [to us] (we speak now of multicellulars), which are similar or different in all degrees. Among them, however, the natural groups of usually particularly similar ones belong together in such a fashion, that they form a reproductive community, i.e., that they can stand in connection with one another, in part as corporal ancestors and descendants, in part as spouses. Such a reproductive community, as long as it is not obtained by artificial force and lasts, i.e., is capable of living through many generations, is called a species.

More than a half century later, Dobzhansky (1970: 23) articulated a model bearing striking similarities to the models of Poulton (1904) and Naef (1919):

In sexually reproducing organisms, the existence of species could in principal, be demonstrated without reference to the discontinuities in their bodily structures, by observing the pairing and procreation of the creatures concerned. Species are more or less discrete reproductive communities. Members of these communities are united by the bonds of sexual unions, of common descent, and of common parenthood....A species is, however, also something else: a supraorganismal zoological system, the perpetuation of which from generation to generation depends on the reproductive bonds between its members.

Most concepts, especially those proposed since the Modern Synthesis, however, have emphasized diverse qualifiers such as reproductive isolation (e.g., Eimer, 1889; Du Rietz, 1930; Dobzhansky, 1937, 1951; Dobzhansky *et al.*, 1977; Mayr 1942, 1957a, b), divergence (e.g., Jordan, 1896; Simpson, 1961; Mayr, 1957a; Nixon and Wheeler, 1990; Frost and Hillis, 1990), ancestor-descendant relationships (Poulton, 1904; Wiley, 1981; De Queiroz, 1998), niche (e.g., Simpson, 1961; Van Valen, 1976; Mayr, 1982:273; Wiley, 1981), fate (e.g., Naef, 1919; Wiley, 1981; Frost and Kluge, 1994), cohesion (e.g., Templeton, 1989; Cohan, 2002), micro-evolutionary forces (e.g., gene flow, genetic drift, and natural selection, e.g., Templeton 1994), reproductive competition (Ghiselin, 1974), or other criteria. Some concepts emphasize operational issues (Cracraft, 1983, 1987; Vrana and Wheeler, 1992; Fitzhugh, 2005). A concept proposed by Johnson *et al.* (1999) is a conflation of select portions from diverse species concepts:

An avian species is a system of populations representing an essentially monophyletic, genetically cohesive, and genealogically concordant lineage of individuals that share a common fertilization system through time and space, represent an independent evolutionary trajectory, and demonstrate essential but not necessarily complete reproductive isolation from other such systems.

Mayr (1957a: 10–11) recognized that his concepts and those of other authors were "secondary, derived concepts, based on underlying philosophical concepts" and he thought that "the analysis of the species problem would be considerably advanced, if we could penetrate through such empirical terms as phenotypic, morphological, genetic, phylogenetic, or biological, to the underlying philosophical concepts." In conflict with that view and the position that we defend here, however, Mayr (1976: 480) also thought that the term species should not be defined in terms of "intrinsic properties."

Ghiselin (1992b) thought it might be fruitful to think about species in terms of "structure" as is the case for models in physics and astronomy. He also described species as "an integrated system, existing at a level above that of the biological individual" (e.g., Ghiselin, 1969, 1997: 15), which highlights the hierarchy of organization that exists between organism and species, a point that we stress. Ultimately, however, he deviated from a purely organizational model like the one we promote and, at the conclusion of his "long argument," proposed that "species are populations within which there is, but between which there is not, sufficient cohesive capacity to preclude indefinite divergence" (Ghiselin, 1997: 305), which emphasizes cohesion, divergence, and fate.

Here, we embrace Ghiselin's (1992b) proposal that thinking about species in terms of structure has merit. The model we defend—*species are networks of organisms integrated by reproductive mechanisms*—emphasizes the intrinsic organization and processes that maintain systems composed of organisms. Most important, the proposed model excludes criteria such as intrinsic reproductive isolation, divergence, and fate, which unnecessarily increase model complexity and result in other problems. In addition, regarding species as networks is in line with the appreciation of the significance of networks in general (e.g., Bornholdt and Schuster, 2003). Networks are frequently noteworthy wherever they occur such as neural networks (*e.g.*, brain) and communications networks (e.g., internet). Networks of organisms, or species, are noteworthy because they are functional units in evolutionary biology and ecology.

**Networks of organisms**

Before proceeding we acknowledge the incompleteness and ambiguity that characterize models of biological systems (e.g, Du Rietz, 1930; Griffiths, 1974; Salthe, 1985, 1993; Hull, 1989; Sober, 1993; Ghiselin, 1997). For example, from an organizational perspective, a unicellular organism (e.g., *Paramecium*) could also be regarded as a free-living cell. We do not agree, however, with Bessey (1908) who argued: "Nature produces individuals and nothing more…Species have no actual existence in nature. They are mental concepts and nothing more…Species have been invented in order that we may refer to a great number of individuals collectively." Rather, we agree with Kinsey (1930) and Griffiths (1974) who argued that species are real in a material way as are individual cells and organisms. That is, networks (and other kinds of lineages) are characterized by material continuity despite apparent trenchant boundaries of individual organisms. Cells and organisms may seem "more real" to humans because their cohesion is not as difficult to comprehend as those of networks.

A fundamental dichotomy among models of various kinds of systems is their hierarchical or non-hierarchical nature (e.g., Griffiths, 1974; Eldredge, 1985; Salthe, 1985, 1993; Zylstra, 1992; Mahner and Bunge, 1997; Ghiselin, 1997). Hierarchical models include those in which lower levels are integrated into higher levels of complexity (= scalar hierarchy of Salthe, 1985, 1993). For example, despite their shortcomings, the following nested models have proved useful for understanding biological systems: cell, tissue, organ, organ system, organism, and ecosystem (Tables 1–2), each of which is characterized by unique intrinsic organization and processes. Non-hierarchical models, such as prokaryote, eukaryote, arthropod, and vertebrate, describe unique organization but are not characterized by nested relationships (Table 2). We think the term *species* is necessary only if there is unique kind of system that requires explanation. As we defend below, *networks of organisms connected by reproduction* represent a kind of system that has properties found nowhere else in nature and associating the term *species* with such structure increases explanatory power in evolutionary biology.

The level of organization between *organism* and *ecosystem* has been the subject of much discussion among evolutionary biologists and systematists. In the framework proposed here, the most general model for this level of organization is *reproduction network or system of organisms integrated by reproductive mechanisms*, which we equate with *species.* Reproduction networks include, for example, conjugation networks and syngamic systems (*sensu* Poulton, 1904, 1908). In conjugation networks, genetic material is transferred directly from one individual to another, rather than being packaged as gametes as they are in syngamic systems. In addition, whereas the products of mating in syngamic systems are new individuals with unique genetic composition, the products of conjugation are pre-existing individuals with reconstituted genetic material—new individuals are not produced until asexual reproduction (e.g., fission) occurs. Notwithstanding their fundamental differences, conjugation networks and syngamic systems can be explained by the more general model, *reproduction network*, or *species.*

Most of the species concept debate has dwelled on one kind of reproduction network, i.e., bisexual species, including Poulton (1904, 1908) who called them syngamic systems. Key points of Poulton's model include the relationship of organisms (one level of organization) to networks (another level of organization). More significant, Poulton's model is one of the earliest of the post-Darwinian era that incorporates an explicit reference to the primary intrinsic process (i.e., syngamy) that integrates such systems. Fertilization (or syngamy *sensu* Hartog, see Poulton, 1908: 60) itself is usually complex and may involve behavior, morphology, or physiology. In addition, diverse meiotic processes facilitate an orderly exchange of genes and increase genetic variation (and presumably survivability) of syngamic systems. Considering the staggering number of eukaryotes, the evolution of syngamy may represent one of the most consequential events in the history of the biosphere; thus, the term *syngamic system* describes an enormous amount of diversity. The significance of syngamy and the networks it produces, however, is usually eclipsed by other issues in most theoretical discussions about species, particularly since the Modern Synthesis. In any case, we think a more general model, i.e., *reproduction network*, is needed to account for the networks formed in prokaryotes and some eukaryotes such as *Paramecium*, as well as bisexual species.

Despite the explicit nature of Poulton's model, he did not describe it as being focused on intrinsic organization as we do. Furthermore, he (1) viewed speciation as a process during which syngamic systems acquire reproductive isolation, (2) recognized subspecies, and (3) found room in his model for "asexual species," all of which are non-trivial deviations from the position that we defend. Thus, Poulton's concept of species is actually more complex than his stated model implies. Such is the case for many other authors. Hennig (1966), for example, repeatedly equated species with the seemingly parsimonious term "reproductive communities," (e.g, Hennig 1966: 65); however, his concept of species is actually more complex and is developed thorough the text (especially Hennig 1966: 28–75). The incongruence between Mayr's species concept and what he practiced was discussed by Frost (2000). The model we proffer, on the other hand, i.e., *species are networks of organisms connected by reproductive mechanisms,* is as simple as it appears. Because it is focused on organization and stripped of contingent properties, the proposed model is less subjective, which, we argue, will increase the rigor of evolutionary studies.

Viewed from the organizational perspective we propose, the units recognized under most species concepts are special classes of reproduction networks. For example, entities recognized under the Biological Species Concept (BSC) are networks that have evolved intrinsic mechanisms preventing reproduction with other systems. Such mechanisms involve intrinsic properties, such as behavior, morphology, or physiology, of organisms and networks, but do not necessarily reflect a pattern of organization shared only by other Biological Species. We tangentially note that hybridization among distantly related species is well documented (e.g., Burkhead and Williams, 1991; Crawford *et al.,* 1993; Jolly *et al.,* 1997; Baird *et al.,* 1998), therefore, the evolution of intrinsic reproductive isolation is not necessarily absolute and strikes us as a less compelling criterion in species models than has been previously argued (e.g., Mayr, 1942). Likewise, divergence, such as that required under the Evolutionary Species Concept, is manifested in diverse ways (e.g., behavioral, molecular, morphological, and physiological evolution) and does not necessarily reflect a pattern of organization shared only by Evolutionary Species. Although the evolution of reproductive isolation and other forms of divergence are significant events in the histories of individual species and stem from the inherent organization of those systems (i.e., isolated reproduction networks evolve because that is the nature of such systems), such events are not relevant in a general model of species. Species recognized using the model crafted here, on the other hand, share a common organizational trait— they are networks of organisms. As such, they are more objectively defined and, consequently, more meaningful comparisons can be made among recognized entities than can be made among more subjectively defined species.

### The population-species paradigm

Most biologists view species as groups of organisms that are composed of less inclusive groups of organisms (e.g., Poulton, 1904; Du Rietz, 1930; Kinsey, 1930; Dobzhansky, 1937; Mayr, 1942; Burma, 1954; Mayr, 1957a, 1957b; Simpson, 1961; Hennig, 1966; Dobzhansky, 1970; Griffiths, 1974; Mayr, 1976; Wiley, 1981; Futuyma, 1986; Avise *et al.,* 1987; Grene, 1987; Atran, 1990; O'Hara, 1993; De Queiroz, 1998; Templeton, 1989; Kluge 1990; Ghiselin, 1992b; Salthe, 1993; Sober, 1993; Frost and Kluge, 1994; Baum and Shaw, 1995; Ghiselin, 1997; Hartl and Clark, 1997; Harrison, 1998; Avise, 2000; Meier and Willmann, 2000; Raven *et al.,* 2002; Bock, 2004; Wiens, 2004; Alleaume-Benharira *et al.,* 2005). A wide assortment of infra-specific groupings have been recognized including *variety* (Darwin, 1859), *avatar* (e.g., Damuth, 1985), *deme* (Gilmour and Gregor, 1939, MacMahon *et al.,* 1978; Eldredge, 1985; Salthe, 1985,1993), *population* (e.g., Simpson, 1961; Emmel, 1976; Mayr, 1976; Futuyma, 1986; Raven *et al.,* 2002), *metapopulation* (e.g., Hanski and Simberloff, 1997; Harrison, 1998), and *subspecies* (e.g., Poulton, 1904; Mayr, 1976). Du Rietz (1930) discussed more than 40 terms for systems composed of organisms including the now obscure *isoreagent, facies, natio, jordanon, konspecies, commiscuum,* and *comparium.* Despite the existing diversity, new terms continue to be coined (e.g., *Distinct Population Segment* [DPS; US Department of the Interior and US Department of Commerce, 1996]; *Management Units* [Mortiz *et al.,* 1995], *Evolutionary Significant Units* [ESU; Ryder, 1986; Waples, 1991; Moritz, 1994, Waples, 1995], *Designatable Units* [DU; Green 2005]; *Least Inclusive Taxonomic Units* [Pleijel and Rouse, 2003]). Here, we reject the longstanding and largely undisputed claim (explicit or implicit) that multiple levels of organization composed of organisms exist in nature. We counter that systems composed of organisms are most parsimoniously explained by a single model and, therefore, a hierarchy of systems composed of organisms cannot exist.

Viewed in terms of intrinsic organization, the wide array of terms associated with systems composed of organisms is characterized by considerable redundancy, which was recognized by Du Rietz (1930) and Kinsey (1937):

Confusion will be avoided, if we call the basic taxonomic unit the species. It is the unit beneath which there are in nature no subdivisions, which maintain themselves for any length of time or over any large area. The unit is variously known among taxonomists as the species, subspecies, variety, *Rasse* or geographic race. It is the unit directly involved in the question of the origin of species, and the entity most often indicated by non-taxonomists when they refer to species. Systematists often introduce confusion into evolutionary discussions by applying the term to some category above the species level.

More recently, Atran (1990, p. 260) concluded "there can be no essential difference between species and variety" and Mayden and Wood (1995) argued the ESU concept might not be needed because ESUs qualify as species under a variety of species concepts (also see Riddle and Hafner 1999). Jonckers (1973) found similar redundancy with regard to the term population. Despite these observations, redundancy among terms associated with systems composed of organisms has remained uncritically tolerated: Gilmour and Heslop-Harrison (1954:151) stated: "Sometimes, however, demological units will coincide with recognised taxa." Simpson (1961, p. 176) claimed a "deme may correspond with a subspecies or even, in extreme cases, with a whole species." Eldredge (1985, p. 152, 153) argued that in some cases demes and populations "may well be virtually coextensive." Kluge (1990, p. 420) stated, "Rarely is a single, uniform deme referred to as a species." De Queiroz (1998) thought that "the population level is really a continuum of levels." That individual systems composed of organisms might be equally regarded as ESUs, demes, populations, race, subspecies, or species, indicates that considerable redundancy is built into the lexicon of evolutionary biologists and systematists (e.g., Table 1). We argue that elimination of such redundancy will lead to clarity in those disciplines.

If organizational distinctions existed among models of systems composed of organisms, it should be possible to illustrate them, as is routinely done with other levels of organization such as atom, molecule, cell, organism, ecosystem, solar system, galaxy, etc. One of the most influential illustrations of systems composed of organisms is that of Hennig (1966: Fig. 6), which has been reproduced repeatedly in this context (e.g., Wiley, 1981; Kluge, 1990; O'Hara, 1993; Graybeal, 1995; and Davis, 1996). Taken at face value, we think Hennig's illustration (Fig. 2) accurately reflects the organization of systems composed of organisms in the natural world and is consistent with the model proposed here. In it, he depicts two levels of organization: (1) organisms (represented by circles) and (2) reproduction networks, which he encircled and identified as species. His illustration also suggests (1) a nearly instantaneous nature of speciation, (2) a narrow zone of "individual fuzziness" at the beginnings and ends of species, and (3) an absence of diagnostic features for each isolated system— all of which are characteristic of the model we promote. Despite the clarity of Hennig's graphical representation, elsewhere he argued in favor of the traditional population-species paradigm (e.g., Hennig, 1966: 47) and he viewed speciation as a process in which "a long time" is required to achieve "complete genetic isolation of successor reproductive communities from an original reproductive community" (Hennig, 1966: 66; see also De Queiroz, 1998), which is contrary to the instantaneous nature of speciation implied by his illustration. In the text, Hennig also used the terms deme, population, and subspecies (e.g., Hennig 1966: 56), which do not appear in his Figure 6, nor is it apparent where they might be incorporated. Other attempts to illustrate the distinction between demes/populations/species fail to depict structurally unique systems. For example, Burma (1954: Fig. 4,) Simpson (1955: Fig. 48), and Kluge (1990: Fig.1) represented species and nested units (e.g., demes, populations) in essentially the same way—as branches of trees. The only distinction between species and less inclusive groups that we are able to infer from such illustrations is that, in some cases, the latter integrate (hybridize) with other lineages and species do not. Such a distinction, however, is not based on the inherent structure of systems but on their fates, which is also problematic (see below). The absence of a graphical representation of the organizational distinction between species and intra-specific units is consistent with our argument that no such distinction exists in nature. We suspect that the proliferation of terms associated with systems composed of organisms reflects, in part, advances in our ability to detect reproduction networks rather than advances in our ability to detect new levels of organization. For example, in many cases, ESUs and MUs represent species that were not evident using morphology alone. If no organizational diversity is represented by the diversity of intra-specific models, obviously, there can be no organizational hierarchy. That which is regarded as a hierarchy of organization is, in many cases, simply a phylogenetic hierarchy among networks.

\*\*\*\*\*\*\*\*\*\*\*\*\*\*\*\*\*\*\*\*\*\*\*\*\*\*\*\*\*\*\*\*\*\*\*\*\*\*\*\*\*\*\*\*\*\*\*\*\*\*\*\*\*\*\*\*\*\*\*\*\*\*\*\*\*\*\*\*\*\*\*\*\*\*\*\*\*\*\*\*\*\*\*\*\*\*\*\*\*\*\*

Figure 2: Hennig's (1966: Fig. 6) representation of a speciation event, which is consistent (when interpreted literally) with the model proposed here. Note the (1) instantaneous nature of speciation, (2) absence of diagnostic feature for each isolated system, (3) narrow zone of "fuzziness" with regard to the origin and extinction of species, (4) emphasis on network structure rather than divergence, intrinsic reproductive isolation, or fate, and (5) the absence of infraspecific terms such as deme and population.

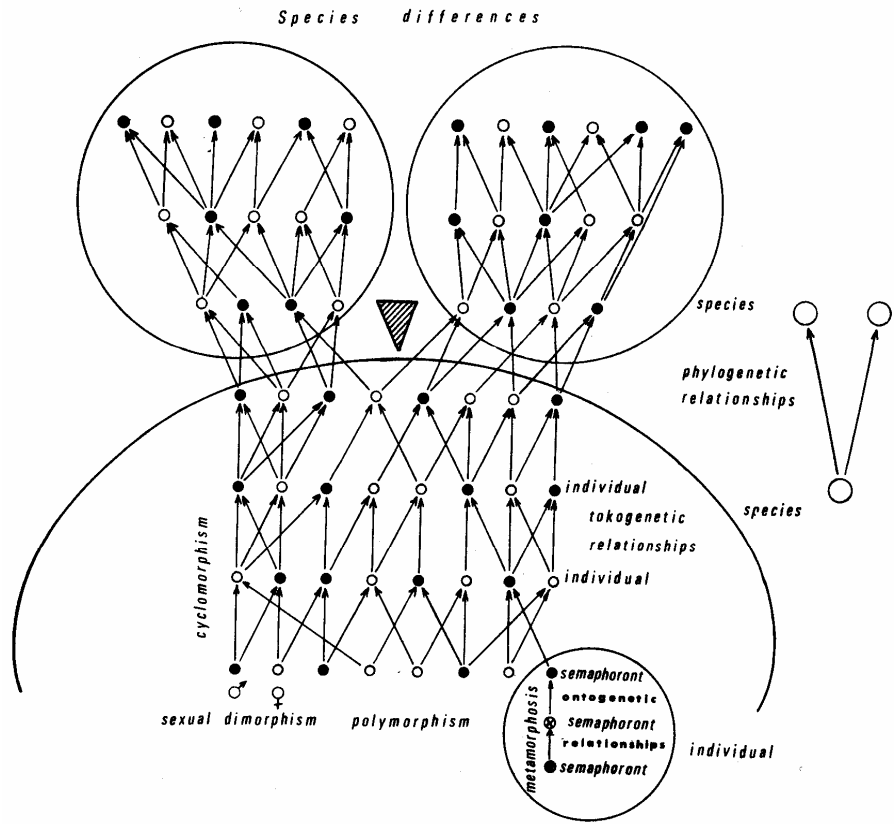

\*\*\*\*\*\*\*\*\*\*\*\*\*\*\*\*\*\*\*\*\*\*\*\*\*\*\*\*\*\*\*\*\*\*\*\*\*\*\*\*\*\*\*\*\*\*\*\*\*\*\*\*\*\*\*\*\*\*\*\*\*\*\*\*\*\*\*\*\*\*\*\*\*\*\*\*\*\*\*\*\*\*\*\*\*\*\*\*\*\*\*

Another indication of the redundancy among the terms used for systems composed of organisms (e.g., deme, population, species) is the absence of unique processes associated with such terms. For example, mutation, meiosis, and syngamy (*sensu* Hartog) occur in populations, demes, and species, regardless of the size or longevity of those systems. In other words, the processes integrating demes, populations, and species are the same, because those systems have the same organization. Gene flow has been cited (e.g., Mayr, 1963; Emmel, 1976; Templeton, 1989; Mallet, 1995) as a *mechanism* that integrates the populations (one proposed level of organization) of species (another proposed level of organization). Gene flow is more accurately viewed, however, as a *consequence* of meiosis and syngamy (*sensu* Hartog) rather a process or "microevolutionary force" (Templeton, 1994) that integrates systems composed of organisms. Stated another way, "genes flow" as a result of the processes that integrate networks of organisms. Because the diversity of terms used for systems composed of organisms (e.g., deme, population, species) are not characterized by unique intrinsic processes, we suspect such terms do not represent unique kinds of systems, much less distinct levels of organization.

Given the lack of transitive shifts (Salthe, 1985) in organization or process among deme/population/species, as seen among other adjacent levels of organization such as atom/molecule/cell/organism, we argue that systems composed of organisms are most parsimoniously explained by a single model rather than multiple models (Table 1, Figs. 2, 3). The lack of attention to organization (and accompanying processes) has resulted in a proliferation of terms associated with systems composed of organisms (e.g., Du Rietz, 1930) without achieving clarity. The species model we promote, on the other hand, is clearly tied to structure and processes, as is the case for other kinds of networks (e.g., Bornholdt and Schuster, 2003) and, we think, will increase understanding of biodiversity.

Under the proposed model, most recognized species likely represent groups of networks, (i.e., species complexes, clades, convenience groupings characterized by plesiomorphic similarity; e.g., Fig 1). Such chunks of diversity will be better understood once they are parsed to reflect the inherent organization of systems rather than contingent properties or other criteria. For example, vexation over "remarkable" intraspecific variation in *Oophaga pumilio* (e.g., Myers and Daly, 1983) will evaporate under the proposed model because much of that variation will be understood to be inter-specific.

In sum, in terms of organization and process, we find nothing to defend the widely held notion that species are composed of smaller networks of organisms (e.g., population, deme). Further, we think the plethora of terms associated with networks of organisms (e.g., Du Rietz, 1930) obfuscates rather than illuminates understanding of biodiversity. Greater clarity can be achieved in evolutionary biology and systematics through implementation of models that explicitly and parsimoniously explain the systems observed in nature. We conclude that a single model, *i.e., network of organisms integrated by reproductive mechanisms*, which we equate with species, sufficiently explains systems composed of organisms.

*********************************************************************************************************

Figure 3: Diagrammatic representation of seven reproduction networks, or species (i.e., Boxes A–G), through time (T1–T6). Circles represent individual organisms, lines represent ancestor-descendant connections, and thick black bars represent vicariance events. Under most species concepts, B, C, E, and F are recognized as temporarily isolated populations of the more inclusive species "H" (dashed box); therefore, two levels of organization are recognized. Here, we argue that systems composed of organisms are represented by a single level of organization (e.g., conjugation networks, syngamic systems) that become smaller and more numerous through vicariance (T2, T4) and larger and less numerous through hybridization (T3, T5). Species originate at the moment they become isolated and become extinct upon integration with other systems.

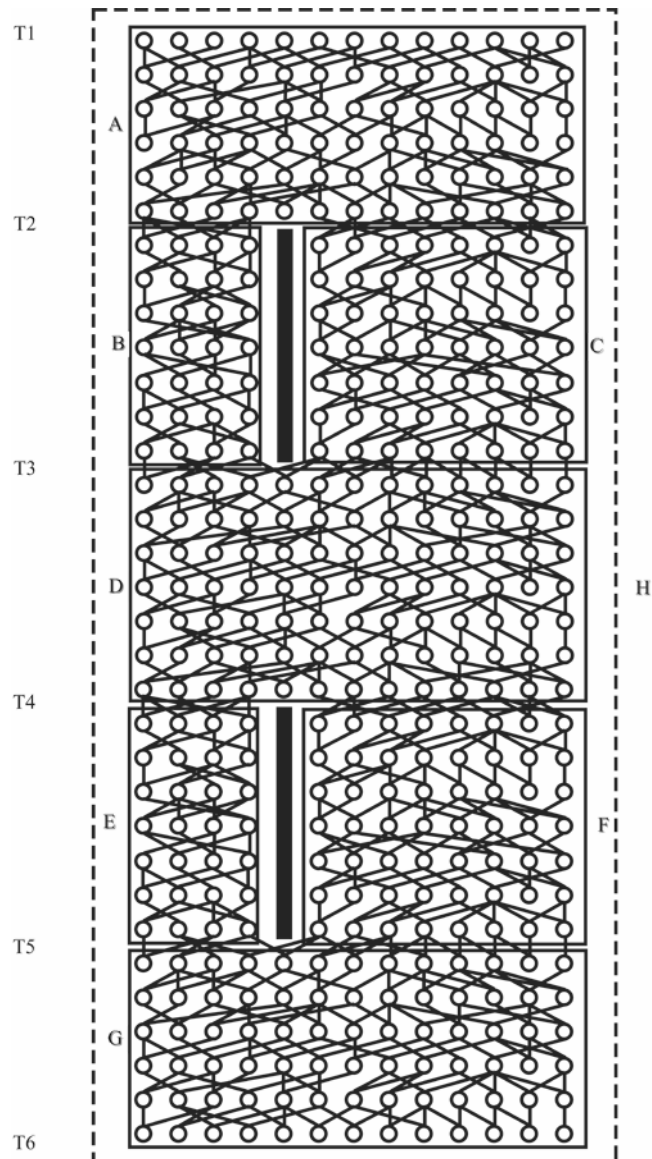

***************************************************************************************************************

**Fate**

Under some species concepts, fate (e.g., permanent isolation, extinction in isolation, lineage reticulation, coalescence with adjacent systems) carries decisive weight in the evaluation of the ontological status of entities. Fate is a conspicuous component of some species models including the Evolutionary Species Concepts (ESC) of Wiley (1978, 1981), Frost and Hillis (1990), Frost and Kluge (1994), Frost (2000), and Chippendale (2000). Under other species concepts (e.g., Jordan, 1896; Dobzhansky 1937, 1970; Hennig, 1966; Baum and Shaw, 1995; Meier and Willmann, 2000; Cohan, 2002), in which species are equated with permanently isolated systems and subspecies or populations are equated with temporarily isolated systems, fate is implied. For example, under most species concepts, isolated networks of nematodes on individual trees are regarded as subunits (e.g., demes, populations) of a species because of the expectation that those networks will integrate *in the future*. Such is the case particularly if individual systems exhibit no divergence. Others (e.g., Darwin, 1838; Naef, 1919; Mayr, 1957b; Hull, 1976; Atran, 1990; Wiens, 2004) require networks to be isolated for some significant (but usually unspecified) period of time before they can be regarded as species. O'Hara (1993) described Wiley's (1978) explicit inclusion of fate in his species concept as "particularly insightful" and he regarded fate as an "ineliminable" characteristic of some concepts because of the "limitations of representation." Below we discuss the problems that result from incorporation of fate in species models and ramifications of its exclusion.

First, under some concepts, it is at least theoretically possible that highly divergent systems, including those that have evolved intrinsic reproductive isolation, will not be recognized as species. For example, if an intrinsic pre-mating isolation mechanism breaks down and allows sister networks to coalesce, under some species concepts (e.g., Mayr, 1957a; Frost and Hillis, 1990; Meier and Willmann, 2000), those systems are considered to be temporarily isolated populations of a species regardless of how much divergence occurred in isolation  (see also Grant, 2002). Therefore, such concepts give more weight to fate than to the intrinsic properties of systems.

Second, although an independent fate may be a function of intrinsic reproductive isolation, it may merely reflect events unrelated to the biology of organisms. For example, if a reproduction network is driven to extinction because of an asteroid impact, the fate of that system will have been determined on the basis of chance external events rather than intrinsic properties that allow or prevent coalescence with its sister system. Nevertheless, under some concepts (e.g., Evolutionary Species Concept), an entity driven to extinction in such a fashion technically qualifies as a species because it remained permanently isolated from all other systems.

The most compelling criticism that can be levied against the inclusion of fate in species concepts is that it is not an inherent property of systems. By definition, systems (e.g., an organized or connected group of objects; OED, 2006) exist by virtue of their organization—regardless of fate. Because fate is a reflection of events involving systems and is not an organizational property or intrinsic feature of systems themselves, we regard it to be an inappropriate consideration in species models. It is relevant to note that fate is absent from models for other levels of organization (e.g., atom, molecule, cell, tissue, organ, organism, ecosystem, star, solar system, galaxy) and other models. For example, trans-actinide elements are recognized because of their unique atomic structure, not their longevity (a few seconds or less) or fate. Likewise, under the model we propose, all reproduction networks have the same ontological status regardless of their longevity or fate (i.e., temporary or permanent isolation; Figs. 1, 3). We think there is no room under the rubric of scientific inquiry for statements about the fate of any such system regardless of how likely reintegration (hybridization) seems. Indeed, it may be just as likely for any particular system to become extinct as it is for it to integrate with other systems.

The elimination of fate from species models would result in the recognition of more biological diversity. It might be argued that implementation of the proposed model will result in recognition of trivial units of diversity and, indeed, a stupefying number of species would be recognized. For example, under the proposed model, every isolated syngamic system of *Ameiva auberi*, lizards which occur on hundreds of islands in the Caribbean, would be recognized as a separate species (Schwartz and Henderson, 1991; see also Fig. 1). Because we emphasize network structure that previously existed or currently exists (rather than structure that could exist or might exist in the future), one possible extension of the proposed model is that multiple isolated networks might exist syntopically.

For example, the American Robins (*Turdus migratorius*) in Central Park might represent multiple species and every tree might harbor a unique species of nematode. The conservation implications of such scenarios are overwhelming to say the least. Most biologists would probably agree with Simpson (1961:151) who argued that "…it would be inconvenient or almost ridiculous to insist that each disjunct population is a separate genetical species" (see also Du Rietz, 1930; Baum and Shaw, 1995; Mahner and Bunge, 1997; Davis, 1996; Bock, 2004; Wiens, 2004). Simpson (1961) did not say, however, why it was ridiculous to do so and we know of no compelling justification for this widely held opinion. Looking again at systems recognized in other disciplines, an atom of molybdenum, a solar system, a kidney, a bacterium is what it is because of its inherent organization, regardless of longevity or fate. Although it might be "inconvenient" to recognize evanescent systems, convenience is not a valid criterion in weighing the merits of scientific models. Moreover, it is not necessary to name every individual reproduction network on Earth—progress is made in astronomy despite the existence of billions of unnamed stars. Regardless, large numbers of species can be easily classified with existing nomenclatural protocols. For example, binomials can be appended with "complex," which draws to attention to species-level diversity without naming individuals systems, e.g., "*Gasterosteus aculeatus* complex" accurately and unambiguously conveys the fact that Three-Spine Sticklebacks are represented by multiple syngamic systems (e.g., Schluter, 1996). Even species complexes with complex histories, such as those in which there have been numerous isolation and coalescence events resulting in a sequence of many networks through time, can be easily reflected in existing classification systems (e.g., Fig. 3).

Given our view of diversity, we expect most currently recognized binomials to correspond to species complexes (e.g., Fig. 1). For example, Macy et al. (2001) argued that the Mountain Yellow-legged Frog (*Rana muscosa*) is composed of multiple "populations." We, however, regard the unambiguous phylogenetic structure and estimated 1.4–2.2 millions of years of isolation hypothesized to characterize those isolated networks as evidence for the existence of at least four species (Kizirian and Donnelly, 2004). Other studies have revealed what we view as species-level diversity at a much finer scale than previously thought. Monsen and Blouin (2004) reported that populations of *Rana cascadae* "show clear isolation by distance" and "a striking pattern of reduced gene flow between populations separated by more than 10 km," Watts et al. (2004) discovered "fine-scale genetic structuring" among isolated networks of damselflies (*Coenagrion mercuriale*) separated by less than 10 km and no evidence of exchange of networks among networks separated by 1000 m. In a classic textbook example, Selander and Kaufman (1975) found evolution among multiple isolated networks of snails (*Helix aspersa*) in two city blocks in Bryan, Texas. Indeed, the trend we observe with recently proposed models (e.g., Distinct Population Segment [DPS; US Department of the Interior and US Department of Commerce, 1996]; Management Units [Mortiz *et al.*, 1995], Evolutionary Significant Units [ESU; Ryder, 1986; Waples, 1991; Moritz, 1994, Waples, 1995], Designatable Units [Green 2005]; Least Inclusive Taxonomic Units [Pleijel and Rouse, 2003]) is recognition of smaller units of diversity. Whereas the criteria for such models reflect some level or type of divergence, however, our model emphasizes inherent organization.

## Some ramifications for speciation and biogeography

In an organizational framework, units of diversity are recognized on the basis of where entities are in time and space, which has two practical manifestations. First, species may be recognized solely on the basis of geographic distribution (e.g., Sober, 2000:156–159; Fitzhugh, 2005) because such information specifies where individual networks are in time and space with respect to other systems (Figs. 1–3). For example, under the proposed model, the Mole Salamander (*Ambystoma talpoideum*) would be recognized as a complex of at least ten species, because it is composed of multiple reproductively disconnected networks isolated by geography (Fig. 1) and despite the absence of features by which those systems could be diagnosed. Arguing that the Mole Salamander represents a single species, on the other hand, appeals to overall similarity, potential for interbreeding, or scientifically unjustifiable predictions about the future (i.e., fate of isolated systems), rather than the inherent organization and histories of those systems. Therefore, whereas most view speciation as a process whereby reproductive communities somehow attain special status, in the organizational framework we adopt here, the mere isolation in time and space of such systems, rather than their evolution, generates species diversity. Indeed, under the proposed framework, the evolution species undergo in isolation is irrelevant in circumscribing them.

Second, under the proposed model, species are recognized at the moment they become isolated (in an analogous fashion, atoms are recognized as such, at the moment they become organized). Consequently, classifications resulting from the implementation of the proposed model will tend to reflect the causal events that generate abundance and diversity of living systems, such as specific vicariance events, rather than events occurring subsequent to isolation, such as character evolution (e.g., Fig. 3). Classifications in the proposed framework will have, therefore, more relevance in the context of evolutionary biology and biogeography in particular. Traditional classifications, on the other hand, generally reflect events occurring after systems have been isolated (e.g., character evolution in isolation) and, consequently, are actually divorced from biogeographical processes.

**Lineage concepts**

As discussed above, species concepts proposed prior to Darwin (1859) tended to emphasize the network component of species systems. Poulton's concept (1904; see above) may have been the first to explicitly include both a network component and a lineage component. Although he abandon it (Simpson, 1961), Simpson (1951) proposed a concept that does not include a network component (i.e., a species is a lineage evolving separately from others and with its own unitary evolutionary role and tendencies). Wiley recycled Simpson's idea (e.g., 1981; a species is a lineage evolving separately from others and with its own historical fate and tendencies) and greatly influenced those who embraced Hennigian principles (e.g., Hennig, 1966). De Queiroz thought that progress could be made by searching for consensus among species concepts and proposed the General Lineage Concept based on the observation that ancestor-descendant relationship is shared among concepts (e.g., de Queiroz, 1998, 2005). Below we discuss problems with lineage concepts that stem largely from a disregard for organization.

Lineage relationships are not unique to species. Indeed, molecules (e.g., DNA), cells, and organisms exhibit lineage relationships. Furthermore, the precursors to life may have been self-replicating, i.e., lineage-forming molecules (e.g., Cech, 1986; Gilbert 1986) and lineages of viruses are regarded by many as non-living (e.g., Raven *et al.*, 2002). Thus, ancestor-descendant relationship is an uninformative criterion in species models because of its generality. De Queiroz (1998; 2005) recognized the generality of lineage relationships but was untroubled by it: "organisms and species (along with genes and cells) are members of the same general category of individuals—lineage-forming biological entities—though they obviously differ with respect to the level of organization." The fact that species are characterized by lineage relationships goes without saying, however, in the same way that it goes without saying that species are characterized by molecular structure.

Focus on ancestor-descendant relationship and disregard for organization results in a more serious problem with Queroz's model. Under the General Lineage Concept (e.g., de Queiroz, 1998, 2005; "species are segments of population level evolutionary entities") the definition of the term *species* is contingent on the definition of the term *population*, which de Queiroz does not define. As we have argued above, species and population cannot be distinguished on the basis of inherent organization (both are networks of organisms), therefore, from our perspective, the General Lineage Concept is characterized by an element of circularity.

One reason for the popularity of lineage concepts may the ease with which they accommodate asexually reproducing organisms, a longstanding issue in the species concept literature (e.g., Poulton, 1904; Frost and Wright, 1988; Fitzhugh, 2005). In asexual lineages, new individuals are formed when parental organisms self-replicate. As such, no supra-organism system is formed and, in general, individuals lack connectivity of any kind (e.g., Sober, 2000: 158). Because they do not form networks, the most complex level of organization represented by asexually reproducing organisms is *organism*, a term that is represented by a relatively trouble-free model (e.g., OED, 2006, but not Raven *et al.,* 2002!). In other words, asexually reproducing organisms require no model to explain them other than *organism* and, therefore, invoking a species model to do so is superfluous. At the same time, however, we agree, in part, with Bessey (1908) that names are useful for discussion of groups of such organisms, as they are for other parts of the tree of life. *Aspidoscelis exsanguis,* for example, is a useful name for certain parthenogenetic lizards occurring in western North America. To reiterate the main point, because asexually reproducing organisms do not form supra-organismal systems (if they indeed do not), a supra-organismal (species) model is not needed to explain them.

In sum, lineage concepts are problematic because of their focus on ancestor-descendant relationship, which is too general to be informative, rather than the unique organization that characterizes systems composed of organisms and the processes that maintain such systems. One of the primary reasons for implementing lineage concepts, classification of asexually reproducing organisms, is argued to be invalid. We conclude that species are best thought of as networks rather than lineages.

## Pluralism, Individuality, and Operational Concerns

We think the most productive way to achieve understanding of biodiversity is to employ models that explicitly and parsimoniously describe the structure of systems in nature. As argued above, systems composed of organisms can be sufficiently explained by a single model, rather than multiple models or concepts of species based on subjective criteria. That is not to say that there are not different kinds of species. As in physics wherein many kinds of atoms are recognized, under our model there are many kinds of reproduction networks including those that have evolved intrinsic reproductive isolation (e.g., BSC) and those that are only weakly divergent from their closest relatives (e.g., ESC). However, whereas in physics understanding of atomic diversity is constrained by a fundamental model based on intrinsic structure, the same cannot be said for species under popular models. Because there is no fundamental model of species based on intrinsic structure, the term species has been applied to many things—including systems spanning multiple levels of biological organization (e.g., de Queiroz, 1998, 2005)—which is convenient but not rigorous practice. Therefore, we reject pluralism (e.g., Kitcher, 1984; Dupre, 1993; Ereshefsky, 1998; Wilson, 2003; Bock 2004) and similar ideas (i.e., homonymy sensu Reydon, 2005) because of the increased model complexity and ambiguity they promote.

Some (e.g., Ghiselin 1969, 1974, 1992a, 1997; Hull 1976, 1989; Sober, 1993) have proposed that species are best thought of as individuals (see also Mayr, 1942: 152). Historical individuals include many things (e.g., Table 2), however, including cell lineages (e.g., *Helacyton gartleri*) and individual organs (e.g., Mickey Mantle's first liver) and, while we do not object to recognition of such diversity, like De Queiroz (1999), we think it is useful to be more specific about the kind of individuals being recognized. Our objective here is to draw attention to one kind of historical individual.

It might be argued that the systems we recognize are too ephemeral and too difficult to circumscribe to be useful. We acknowledge that there are temporal and spatial issues that might make identification of networks difficult. Operational challenges, however, are insufficient cause for rejecting theoretical models and, in any case, other species concepts or models are not universally easier to implement than the one we propose. In some cases identification of species will actually be easier in the proposed framework because all that is needed is a distribution map (e.g., Fig. 1). In any case, we think the theoretical issues characteristic of widely implemented species concepts are of greater concern than the operational issues posed by the model proposed here.

## Summary

Viewing diversity in terms of inherent organization leads to several interrelated conclusions. We reject the population-species paradigm with its diverse models (e.g., Evolutionary Significant Unit, deme, population, species, etc.) because none can be distinguished on the basis of unique structure or process. In other words, systems composed of organisms are most parsimoniously explained by a single model rather than multiple models. Temporarily isolated systems, typically unrecognized under current concepts of species, are organizationally and, therefore, ontologically indistinguishable from permanently isolated systems. Fate is an inappropriate consideration in species concepts because it is not an intrinsic property of systems and is notably absence from models in other scientific disciplines. Lineage concepts are problematical because ancestor-descendant relationship is too general a criterion to be informative at the species level and because such concepts disregard the organization of systems composed of organisms.

Under the new view of diversity proposed here, the "furniture of the biosphere" consists of (1) organisms and (2) networks of organisms, the latter equated with "species". The proposed model, i.e., *species are networks of organisms integrated by reproductive mechanisms*, is characterized by compelling features, including unique organization (i.e., network of organisms) and unique intrinsic processes (e.g., conjugation, meiosis, and syngamy). Classifications reflecting the proposed model better reflect biogeographic processes because names are tied to the events that isolate networks of organisms rather than the subsequent evolution of those networks. The proposed model also results in greater unity of thought among scientific disciplines because, like many models in other fields, it is focused on the inherent organization of systems. As a final note, the total number of species under the proposed model is considerably higher than the previous estimates (e.g., May, 1988; Wilson, 2000) and closer to those of other levels of organization. That is, atoms, molecules, cells, organisms, stars, galaxies—and, we think, species—number in the billions. The picture that emerges from implementation of the proposed model might seem overwhelming in some respects; however, we think it better reflects the dynamic nature and actual diversity of biological systems.

## Acknowledgements

We thank Tom Amorosi, Kimberlee Arce, Sumru Aricanli, Margaret Arnold, Steve Ashe, Julian Avery, Raoul Bain, Kristen Bell, Brian Brown, Don Buth, Lisa Campbell, Jim Carpenter, Alessandro Catenazzi, Catherine Childs, Jack Conrad, Brian Crother, Brian Danforth, David Dickey, Adam Edwards, Dilek Erdal, Bobby Espinoza, Kirk Fitzhugh, Javier Francisco-Ortega, Darrel Frost, Sherri Gabbert, Kimball Garrett, Dan Geiger, Michael Ghiselin, Andy Gluesenkamp, Taran Grant, Craig Guyer, Mike Hardman, Michael Harvey, Gordon Hendler, Ines Horovitz, Jerry Johnson, Shari Kizirian, Arnold Kluge, Michel Laurin, Rich Leschen, Carl Lieb, Darrin Lunde, Jody Martin, Kendall Martin, Dan Meinhardt, Art Metcalf, Bob Minckley, Adrian Nieto, Steve Oberbauer, Richard Pearson, Chris Raxworthy, Brian Rosen, Thomas Reydon, Kia Ruiz, Ralph Saporito, Jay Savage, Chris Thacker, Kevin Toal, Dick Vane-Wright, Stephen Vail, Leigh Van Valen, Emilie Verdon, Rudolf von May, Xiaoming Wang, James Watling, Bob Webb, Steven Whitfield, Erik Wild, Ed Wiley, P. Wilson, and John Wright for their input, W. M. Keck Foundation and the R. M. Parsons Foundation for funding, and Dan Geiger for translating portions of Naef (1919). This is Contribution No. 151 of the Tropical Biology Program at Florida International University.